\documentclass[11pt]{article}
\usepackage{jheppub}

\usepackage{amsmath}
\usepackage{latexsym,amssymb}
\usepackage{graphicx,color,slashed}
\usepackage{bbm} 
\usepackage{ifpdf}

\newcommand{\cG}{\mathcal{G}}

\newcommand{\cH}{\mathcal{H}}
\newcommand{\cL}{\mathcal{L}}

\newcommand{\cN}{\mathcal{N}}

\newcommand{\Tr}{\mathrm{Tr\,}}

\newcommand{\be}{\begin{equation}}
\newcommand{\ee}{\end{equation}}
\newcommand{\ba}{\begin{eqnarray}}
\newcommand{\ea}{\end{eqnarray}}

\renewcommand{\d}{\textrm{d}}
\newcommand{\e}{\textrm{e}}

\renewcommand{\a}{\alpha}
\renewcommand{\b}{\beta}
\newcommand{\N}{\mathcal{N}}

\def\E{{$E_{7(7)}$}}

\def\La {{\Lambda}}

\def\g{\gamma}
\def\d{\delta}

\def\e{\hat e}

\newcommand{\rf}[1]{(\ref{#1})}
\newcommand{\bea}{\begin{eqnarray}}
\newcommand{\eea}{\end{eqnarray}}

\def\bfzero{\relax{\rm I\kern-.18em 0}}
\def\bfone{\relax{\rm 1\kern-.35em 1}}
\def\twomat#1#2#3#4{\left(\begin{array}{cc}
\end{array}
\right)}

\newcommand{\mc}[1]{\mathcal{#1}}

\newcommand{\D}{\Delta}

\newcommand{\cn}{{\cal N}}
\def\pa{\partial} 
\def\>{\rangle} 
\def\<{\langle} 
\def\k{\kappa}

\def\Tr{{\rm Tr}}

\def\a{\alpha}
\def\b{\beta}

\def\d{\delta}
\def\e{\epsilon}           
  
\def\g{\gamma}

\def\k{\kappa}                    
\def\l{\lambda}

\def\D{\Delta}

\def\pa{\partial}                              
\def\>{\rangle} 
\def\<{\langle} 
\def\Dsl{D \hskip-.6em \raise1pt\hbox{$ / $ } }
\def\to{\rightarrow}
\def\pa{\partial}

\newcommand{\lra}{\leftrightarrow}

\def\da{{\dot\alpha}}
\def\db{{\dot\beta}}

\def\tQ{\tilde{Q}} 
\newcommand{\reef}[1]{(\ref{#1})}

\newcommand{\Si}{\Sigma}

\title{Duality Constraints on Counterterms in $\cn=5,\,6$ Supergravities}

\author[a,b]{Daniel Z. Freedman,}
\author[a]{Renata Kallosh,}
\author[a]{and Yusuke Yamada}
\affiliation[a]{Stanford Institute for Theoretical Physics and Department of Physics, Stanford University, Stanford, CA 94305, USA}
\affiliation[b]{Center for Theoretical Physics and Department of Mathematics, Massachusetts Institute of Technology, Cambridge, MA 02139, USA}\emailAdd{dzf@math.mit.edu}
\emailAdd{kallosh@stanford.edu}
\emailAdd{yusukeyy@stanford.edu}

\notoc

\abstract{The UV finiteness found in calculations of the 4-point amplitude in  $\cN=5$ supergravity at loop order $L=3, 4$ has not been explained, which motivates our study of the relevant superspace invariants and on-shell superamplitudes
for both $\cN=5$ and $\N=6$. 
The local 4-point superinvariants for $L = 3,4$ are expected to have nonlinear completions whose 6-point amplitudes have non-vanishing SSL's (soft scalar limits), violating the behavior required of Goldstone bosons.
For $\cn=5$, we find at $L=3$ that local 6-point superinvariant and superamplitudes, which might cancel these SSL's, do not exist. This rules out the candidate 4-point counterterm and thus
gives a plausible explanation of the observed $L=3$ finiteness.  However, at $L= 4$ we construct a  local 6-point superinvariant with non-vanishing SSL's,   so the SSL argument does not explain the observed $L=4$ $\cN=5$ UV finiteness. For $\cN=6$ supergravity there are no 6-point invariants at either $L= 3$ or 4, so the SSL argument predicts UV finiteness. 

}

\begin{document}

\maketitle

\newpage

 \tableofcontents{}


\section{Introduction}
The ultraviolet properties of supergravity theories in 4 spacetime dimensions have been explored since the theories were first discovered in 1976.
Most attention has focused on the maximal $\cN=8$ theory  \cite{Cremmer:1979up}, \cite{deWit:1982bul}.  Candidate counterterms  (CT) were proposed long  ago \cite{Kallosh:1980fi,Howe:1980th,Howe:1981xy}.
A series of difficult, intricate calculations has shown that the 4-point graviton scattering amplitude is finite at loop level $L=3,\,4$  \cite{Bern:2007hh,Bern:2009kd}. Results suggested that the critical dimension at which ultraviolet divergences first occur is the same as that of the $\cN=4$ SYM theory, namely $ D= 4 + 6/L$. Recently, however, a  5-loop calculation \cite{Bern:2018jmv} indicates that the amplitude first diverges at critical dimension  $D=24/5$.  Theoretical approaches based on both linearized superspace invariants~\cite{Drummond:2003ex,Drummond:2010fp} and local superamplitudes~\cite{Elvang:2010jv,Elvang:2010kc,Beisert:2010jx} agree that the conventional symmetries of supergravity allow candidate counterterms (in integer dimension $D=4$) beginning at $L=7$ loops. One must hope for new mechanisms of enhanced cancellation which apply beyond the 7-loop barrier if the theory is to be finite to all orders.

Although calculational techniques have vastly improved in the course of the program which led to the 5-loop work, 7 loops is a formidable challenge.  Therefore the $\cN=4$ and $\cN=5$ supergravities have been studied. These theories are expected to diverge first at loop level $L=\cN-1$, so one has the opportunity to
search for unexpected cancellations in a simpler setting than in $\cN=8$. The calculations in  \cite{Bern:2014sna} have shown that the  4-point superamplitude in $\cN=5$ supergravity \cite{Cremmer:1980gs},  \cite{deWit:1981yv} is finite at 3- and 4-loop order.   It was also recently established in \cite{Freedman:2017zgq}, that $\cN \geq 5$ models, as opposite to  $\cN< 5$  \cite{Carrasco:2013ypa}, do not have 1-loop anomalous amplitudes. The question is whether conventional symmetries explain these results or whether there are new mechanisms at work.

The unexplained UV finiteness in $L=3,4$ $\cN=5$ supergravity stimulated a recent study of whether it is possible to deform the classical action to compensate UV divergences \cite{Kallosh:2018mlw}.
The analysis was found to be consistent with a restricted version of duality symmetry taken at the base point of the moduli space, but it was difficult to establish that  the deformed action has a supersymmetric embedding. In this sense the analysis in \cite{Kallosh:2018mlw} is inconclusive.

In this paper we investigate to what extent the UV finiteness in $L=3,4$ $\cN=5$ supergravity can be  explained by the soft scalar limit of amplitudes produced by candidate CTs. This approach was very useful
for $\cN=8$ supergravity
\cite{Brodel:2009hu,Elvang:2010kc,Beisert:2010jx,Elvang:2010xn}.  It
incorporates as an essential element the fact that the scalar fields of extended supergravities are the coordinates of a ${\cG\over \cH}$ manifold ($E_{7(7)}/SU(8)$ in the $\cN=8$ theory and $SU(5,1)/(SU(5)\times U(1))$ for $\cN=5$). 
The scalars are Goldstone bosons of the non-linearly realized symmetry, so amplitudes must vanish as the momentum  of any scalar approaches $p^\mu\to 0$. Restrictions on allowed counterterms are obtained by enforcing this requirement.

These restrictions are obtained as follows.  The $\cn=5,6,8$ supergravity theories each possess $\pa^{2k}R^4$ supersymmetric invariants\footnote{Their superfield description is discussed in Sec.~\ref{CTcandidates} below. Exceptionally for $\cn=8,~k=1$, the 4-loop invariant vanishes; its would-be contribution to the scattering amplitude carries the factor $s+t+u=0$.}  for all $k \geq 0$.  These are candidate counterterms for the graviton scattering amplitude at loop level $L = k+3$. If the divergence represented by a counterterm is actually present in the amplitude, the counterterm must be added to the Lagrangian in a supersymmetric fashion.

Once added, the counterterm induces a
 nonlinear and non-local supersymmetric completion. This is very difficult to calculate using field theory techniques, but in the $\cn=8$ theory,  the answer for the soft limit of its  NMHV 4-graviton, 2-scalar matrix elements $\<\bar\phi\,\phi ++--\>$  was extracted from tree level closed string theory amplitudes at 3-loop \cite{Brodel:2009hu,Elvang:2010kc}  and then 5- and 6-loop \cite{Elvang:2010kc,Beisert:2010jx,Elvang:2010xn} order.  Their soft limits do not vanish, and there are no $\phi^2\pa^{2k}R^4$ counterterms which might cancel them, so the $k=0,\,2,\,3$ quartic invariants are ruled out. There are
6-point invariants with non-vanishing soft limits beginning at 7-loop level, so this line of argument suggests that $\cn=8$ supergravity might diverge beginning at this order.

In this paper we are concerned with the implications of soft scalar limits (SSLs) for the ultraviolet behavior of $\cn=5,\,\,6$ supergravity.  String theory cannot be applied to obtain the soft limits of the supersymmetric completion of the quartic invariants in these theories, but it is reasonable to assume that they do not vanish.  So our principal goal is to study 6-point NMHV superspace invariants and the corresponding superamplitudes that contain $\phi^2 R^4$ component amplitudes which potentially cancel the soft limits.  
An  understanding of UV finiteness of  $\cN=5$ at $L=4$ might shed some light on what to expect in the  
related case of $\cN=8$ at $L=7$; in both theories we study the case of $L=\cN-1$ loop order.

Our 6-point invariants are constructed as full superspace integrals of dimensionless superfields $W(x,\theta,\bar\theta)$  \cite{Kallosh:1980fi,Howe:1980th,Howe:1981xy},
whose lowest components are scalar fields $\phi(x)$.  The $W$ superfields are connected
by a chain of Bianchi identities to chiral superfields recently studied in \cite{Freedman:2017zgq}.
Chiral superfields are not directly relevant to the question of non-vanishing SSLs since they contain scalar
fields covered by derivatives. Thus soft limits vanish.

One feature of our work is the emphasis on the correspondence between superspace invariants and superamplitudes.  Both approaches incorporate the linearized (i.e. free field) SUSY transformations between the component particle states of the theory.  At the operational level they look very different, yet they produce the same physical results.  

In Sec.~\ref{CTcandidates} we discuss the 4-point superspace invariants of $\cn=5,6,8$ supergravities at loop level $L=3, 4$. 
Although fully supersymmetric, they are expressed as integrals over suitably restricted subsets of the full superspace of  $4\cn\,\theta$'s. In Sec.~\ref{NMHVsix} we describe the general structure of the  6-point $L \geq \cN-1$ superspace invariants with full details of the $\theta$-expansion in the case $\cn=5,\, L=4$.  In Sec.~\ref{duality} we discuss the duality symmetries of $\cn \geq  5$ supergravity.  These dualities act via shift symmetries of the scalar fields of the theories, and we extend these symmetries to superfields.  The 6-point invariants violate shift symmetry indicating that their scalar amplitudes have non-vanishing SSLs. In Sec.~\ref{N5SA} we discuss superamplitudes for $\cn=5$ and $\cn=6$. A brief summary follows in Sec.~\ref{summary}. Further details on superampliutudes are in Appendices \ref{NMHVgeneral} and \ref{examples}.

Let's state our main results concerning UV divergences.  

\noindent The $\cn=5$ supergravity has no 6-point 3-loop counterterm so the expected non-vanishing SSL from the nonlinear completion of its $R^4$ invariant cannot be cancelled.  This gives a post hoc explanation of the observed finiteness.  

At the 4-loop level there are both $\pa^2 R^4$  and $\phi^2R^4$ invariants, the latter with non-vanishing SSL.  We cannot say whether this SSL cancels between the 6-point invariant and the nonlinear completion of the 4-point. Therefore an  explanation of  $\cN=5$, $L=4$ UV finiteness, discovered in \cite{Bern:2014sna}, is still absent.

In the $\cn=6$ theory there are 4-point candidate UV divergences starting from $L\geq 3$. The 6-point 3- and 4-loop candidate counterterms are absent, they start at $L=5$ level. Thus we predict UV finiteness for both $L=3,4$.  At $\cN=6$, $L=5$  the situation is  unclear,  as it is in $\cN=8$, $L=7$ and $\cN=5$, $L=4$.

\section{Candidate 4-point Counterterms  in $\cn\geq 5$ Supergravities}\label{CTcandidates}
 
All linearized chiral superfields of $\cn= 5,\,6,\,8$ supergravities of dimension $1/2, 1, 2$ are described in \cite{Freedman:2017zgq}.  Their scalar components are covered by derivatives which means that invariants formed from them have vanishing SSL's.  In this paper we feature  linearized superfields of dimension $0$, generically called $W(x,\theta,\bar\theta)$.
Their lowest $\theta$ components are scalar fields without derivatives. Therefore the corresponding superspace invariants may have  non-vanishing SSL's.  The $W$ fields are not  chiral, but they are related to the chiral superfields by chains of Bianchi identities given in \cite{Freedman:2017zgq}. 

In $\cN=8$ we use the superfield $
W^{abcd}(x, \theta)$ which satisfies the conditions derived in \cite{Howe:1981gz}. Here we use notation in \cite{Freedman:2017zgq}
\be\label{N8sf}
W^{abcd}= {1\over 4!} \epsilon^{abcdefgh} W_{efgh},
\ee
\be
D_{\alpha }{}^a W_{bcde}=4\delta^a_{[b}\chi_{cde]\alpha},
\ee
\be
\bar D_{\dot \alpha a}  W_{bcde}=\frac{1}{3!}\epsilon_{abcdefgh}\bar\chi_{\dot \alpha}^ {fgh}.
\ee
In $\cN=6$ we work with $W_{ab78}= W_{ab}$ which satisfies
\be
W_{ab}= {1\over 4!} \epsilon_{abcdef} W^{cdef},
\ee
\be
D_{\alpha }{}^a  W_{bc}=2 \delta_{[b}^{a} \chi_{ c]\alpha},
\ee
\be
D_{\dot \alpha a}  W_{bc}= \frac{1}{3!}\epsilon_{abcdef}\bar{\lambda}_{\dot\alpha}^{def}.
\ee
Finally, in $\cN=5$ we have a superfield $
W_{a678}= W_{a} $ 
that  satisfies the conditions
\be
W^{a}= {1\over 4!} \epsilon^{abcde} W_{bcde},
\label{n51}\ee
\be
D_{\alpha }{}^a  W_{b}=\delta^a_{b} \chi_\alpha,
\label{n52}\ee
\be
\bar{D}_{\dot \alpha a}  W_{b}= -\frac{1}{3!}\epsilon_{abcde}\bar\lambda_{\dot \alpha}^{cde}.
\label{n53}\ee
The $D_\a$ and $D_{\da}$ conditions above are part of the chain of Bianchi identities of \cite{Freedman:2017zgq}.

\subsection{3-loop 4-point  local superinvariants}

The 3-loop 4-point candidate UV divergences furnish a supersymmetric version of the $R^4$ CT. For $\cN=8$ it was proposed in \cite{Kallosh:1980fi} in a form which was manifestly supersymmetric but not manifestly $SU(8)$ invariant; it was only $SU_1(4)\times SU_2(4)$ invariant. We have split $a={I=1,2,3,4, \,J=5,6,7,8}$.  It was observed there that there is a basis $\hat x(x, \theta_I, \bar \theta^J)$  where the superfield with lower indices taking values in $I$  is equal to the superfield with  upper indices taking values in $J$, and depends only on 16 Grassmann variables $\theta_{I=1,2,3,4}$
and on $\bar \theta^{J=5,6,7,8}$
\be W_{1234} (\hat x, \theta_I, \bar \theta^J)=W^{5678} (\hat x, \theta_I, \bar \theta^J) \equiv W (\hat x, \theta_I, \bar \theta^J).
\ee 
 Therefore the integral over these 16 $\theta$'s is a superinvariant
\be
CT_{L=3}^{\cN=8}=\kappa^4 \int d^4 x \, d^{8}\theta_I  d^{8}\bar \theta^J  \,  W^4 (\hat x, \theta_I, \bar \theta^J)= \kappa^4 \int d^4 x  \,  C_{ \alpha  \beta  \gamma  \delta}(x)
 C^{ \alpha  \beta  \gamma  \delta}(x)
  \bar C_{\dot \alpha \dot \beta \dot \gamma \dot \delta}(x)
\bar C^{\dot \alpha \dot \beta \dot \gamma \dot \delta}(x) +\cdots.
\label{3loop}\ee
Here $C_{ \alpha  \beta  \gamma  \delta}(x)$ is the symmetric multi-spinor form of the Weyl tensor.

An improved version of this CT with manifest $SU(8)$ symmetry was proposed in~\cite{Howe:1981xy}.  The expression in~\cite{Howe:1981xy} was given
 in a form universal for all $\cN$,  where the measure of integration as well as the kernel are both  $SU(\cN)$ tensors. The kernel depends on a spinor superfield $ \chi^\alpha _{abc} (x, \theta, \bar \theta)$ whose first component field is a  spin 1/2 field, $\chi _{abc}^\alpha(x)$.  
The universal 3-loop candidate CT \cite{Howe:1981xy} is
 \be
CT_{L=3}^{\cN\geq 4}= \kappa^4 \int d^4 x \, D^{a_1 b_1, a_2, b_2, a_3, b_3} \bar D_{c_1, d_1, c_2, d_2, c_3, d_3} K_{a_1 b_1, a_2, b_2, a_3, b_3}^{c_1, d_1, c_2, d_2, c_3, d_3}
\label{HST}\ee
where  the kernel  is a product of 4 superfields of dimension 1/2. 
\be
 K_{a_1 b_1, a_2, b_2, a_3, b_3}^{c_1, d_1, c_2, d_2, c_3, d_3}
=  \chi _{a_1a_2a_3}^\alpha (x, \theta, \bar \theta)  \chi _{b_1 b_2 b_3 \alpha} (x, \theta, \bar \theta)  {\bar \chi }^{c_1 c_2 c_3}_{\dot \alpha} (x, \theta, \bar \theta)  {\bar \chi }^{d_1 d_2 d_3 \dot \alpha} (x, \theta, \bar \theta)+ {\rm symmetrizations}.
\label{HST1}\ee
This expression has  manifest $SU(\cN)$ invariance\footnote{See \cite{Bossard:2011tq}  for 4-point counterterms derived from harmonic superspace.}. To prove its supersymmetry requires  use of  the Bianchi Identities shown in the previous section, starting with \rf{N8sf}.
The measure of integration for any $\cN$ has dimension $-4+6$, and the kernel has dimension $+2$.  Multiplication  by $\kappa^4$ produces a dimensionless supersymmetric and $SU(\cN)$ invariant which we identify as a 3-loop CT for all $\cN$. For $\cn=8$, this form of the counterterm agrees with \reef{3loop}.

An attempt  to use the same construction to produce a  6-point local 3-loop CT fails. This can be seen follows, in $\cN=8$ case. The 6-point superinvariant of the required dimension can only depend on same type of a superfield, which depends on $\theta_{I=1,2,3,4}$
and on $\bar \theta^{J=5,6,7,8}$ and has the form  $
CT_{L=3}^{\cN=8}=\kappa^4 \int d^4 x \, d^{8}\theta_I  d^{8}\bar \theta^J  \,  (W (\hat x, \theta_I, \bar \theta^J))^6$. The first term here is now given by the following expression, 
$ \kappa^4 \int d^4 x  C^2 \bar C^2 (W_{1234})^2$. Thus, the 6-point superinvariant breaks $SU(\cN)$.
Alternatively, one  finds that a proof of supersymmetry in  \cite{Howe:1981xy} for the manifestly invariant 4-point counterterm does not extend to 6-point generalizations.

\subsection{4-loop  4-point  local superinvariants}
To make a 4-loop candidate which is a supersymmetric extension of $\pa^2R^4$, we can start with the 3-loop form in \rf{HST}  and raise the dimension of the integrand. This means  that we insert two space-time  derivatives acting on the superfields.   In the $\cN=8$ case it is easier to visualize in the form \rf{3loop} where we obtain
\be
CT_{L=4}^{\cN=8}= \kappa^4 \int d^4 x \, d^{8}\theta_I  d^{8}\bar \theta^J  \,  \Big (\partial_\mu  W^2 (\hat x, \theta_I, \bar \theta^J)\Big )^2  \, .
\ee
 This vanishes on shell, as is easy to see  in  momentum space, where
\be
CT_{L=4}^{\cN=8}=\kappa^6  \int  d^{8}\theta_I  d^{8}\bar \theta^J\, (s+t+u) W(p_1, \theta_I, \bar \theta^J)  W(p_2,\theta_I, \bar \theta^J)   W(p_3,\theta_I, \bar \theta^J)   W(p_4,\theta_I, \bar \theta^J) =0.
\label{stueffect}\ee
In $\cN=5$ the candidate $L=4$ CT was given \cite{Freedman:2017zgq} as a chiral superspace integral
\be
CT_{L=4}^{\cN=5}=\kappa^6 \int d^4 x \, d^{10} \theta \, \bar \chi^{\dot \alpha}(x, \theta) \bar \chi_{\dot \alpha}(x, \theta) \bar C_{\dot \alpha \dot \beta \dot \gamma \dot \delta}(x, \theta)
\bar C^{\dot \alpha \dot \beta \dot \gamma \dot \delta}(x, \theta).
\label{candidate}\ee
This integrand has a very different structure, not of the form $(\pa W^2)^2$, so the previous $s+t+u$  argument does not apply to $\cN=5$. The candidate CT in eq. \rf{candidate} remains a candidate for the 4-loop UV divergence.  

One might try  to produce a 4-loop counterterm by raising the universal form of the 3-loop counterterm in \rf{HST}, \rf{HST1}. However the integrand of \reef{HST1} contains the product of two $\chi_\a$ and two $\bar \chi_{\da}$ superfields, so the $stu$ symmetry cannot be used for any value of $\cn.$ 

At loop level $L=4$ in $\cn=6$ supergravity, the  local $\pa^2R^4 $ counterterm can be given by an insertion of  two derivatives in  the $\cN=6$ version of expression in \rf{HST1}. 
We will see later, in Sec. 5  that there is also a related  superampliutde.

 \section{6-point $L\geq \cN -1$ local NMHV superinvariants in $\cn\geq 5$ supergravities}\label{NMHVsix}
 
 The 5-point local superinvariants, start from $L=\cN$ loop order, i.e. $L=5,\, \cN=5$ and  $L=6, \,\cN=6$ and  $L=8,\, \cN=8$, and their SSL's vanish.  Therefore, for the purpose of our investigation we proceed directly to  6-point candidate CT's, which start at order $L=\cn-1$ and have non-vanishing SSL's.

An $L$-loop full superspace counterterm in $\cn$-extended supergravity takes the form:
\be
S_{CT} = \k^{2(L-1)} \int d^4x \,d^{2\cn}\theta\, d^{2\cn}\bar\theta\, \cL_{[2( L+1 - \cn)]}(x,\theta,\bar\theta).
\ee \label{sctgen}
In general, the  Lagrangian $\cL_{[2( L+1 - \cn)]}$ is a (composite) \emph{local} superfield of mass dimension $2(L+1-\cn)$.  Locality 
requires non-negative dimension, hence $L \geq \cN-1$.
We look first at the case $L=\cn-1$ in which $\cL_{[2( L+1 - \cn)]}= \cL_{[0]}$ is \emph{dimensionless}.  At lower loop level a full superspace invariant would be non-local. For example, no local 6-point $L=3$, $\cN=5$  superinvariants are available.

Each $\cL_0$  is a 6th order product of dimensionless elementary superfields whose lowest $\theta$ components are the scalar fields of the theory.  Indeed, these are the antisymmetric $SU(\cn)$ tensors of rank $\cn-4$ described in Sec.~\ref{CTcandidates}.
We deal throughout this paper with linearized on-shell superfields. Their $\theta$ expansions encode the SUSY transformation rules of the on-shell components of the \emph{free} fields of the theory.  These are defined as spinor and symmetric multispinor spinors, e.g.  $\chi_\a,\,\bar\chi_\da$ for helicity $\pm\frac12$ and $F_{\a\b}, \bar{F}_{\da\db}$ for helicity $\pm 1$, etc.  

In $\cN=8$ for the 6-point superamplitudes we have 
\be
\kappa^{12} \int d^{32}\, \theta d^4 x \,   \Tr \Big ((W \bar W)^3 \Big )  \, ,  \qquad \kappa^{12} \int d^{32}\, \theta d^4 x \, ( \Tr W \bar W)^3
\label{N8}\ee
which has a 4-graviton-2-scalar amplitude, a partner of $D^4R^6$. The corresponding NMHV 6-point local manifestly supersymmetric amplitudes are given in \cite{Beisert:2010jx,Elvang:2010xn}.

In $\cN=6$ we have
\be
\kappa^{8} \int d^{24}\, \theta d^4 x \,   \Tr \Big ((W \bar W)^3 \Big ) \, ,  \qquad \kappa^{8} \int d^{24}\, \theta d^4 x \, ( \Tr W \bar W)^3
\label{N6}\ee
which has a 4-graviton-2-scalar amplitude, a partner of $ R^6$.

In $\cN=5$ there is a single  local 6-point invariant 
\bea
\kappa^{6} \int d^{20} \theta \, d^4 x \,  ( \Tr W \bar W)^3 \, \label{5'} .
\label{N5}\eea
There is no 6-graviton amplitude here, as opposed to $\cN=8,\,6$. The reason is that in the scalar superfield $W_a$, the $C,\, \bar C$ multi-spinors appear multiplied by at least 4 $\theta$'s.  Therefore 6 gravitons require 24 $\theta$'s in the measure of integration. These are available in $\cN=8,\,6$ but not in $\cN=5$, where we have only 20 $\theta$'s. Another way to see this is by dimension counting; there is no local term  $ \int d^4x R^6$ at the $\kappa^6$ order (nor is there a local $\psi^2R^4$ component, since the gravitino field $\psi_{\a\b\g}$ has dimension 3/2).

The invariant \reef{N5} contains the 2-scalar, 4-graviton amplitude
\bea
  \kappa^{6}  \int d^4 x \, \partial_{\epsilon\dot\epsilon}\Big( \phi_a (x)
C_{ \alpha  \beta  \gamma  \delta}(x)
 C^{ \alpha  \beta  \gamma  \delta}(x)\Big)
\partial^{\epsilon\dot\epsilon} \Big (\bar \phi^a (x)  \bar C_{\dot \alpha \dot \beta \dot \gamma \dot \delta}(x)
\bar C^{\dot \alpha \dot \beta \dot \gamma \dot \delta}(x)\Big ) +\cdots.
\label{2sc4grA}\eea
The specific deployment of derivatives is taken from the corresponding NMHV superamplitude in Sec.~\ref{N5SA}. 
The 6-scalar component amplitude is 
\bea
  \kappa^{6}  \int d^4 x \,\big(\phi_{(a }(x)\phi_b (x)\phi_{c)} (x)\big) \Box^5\big(
  \bar \phi^a (x) \bar \phi^b (x)  \bar \phi^c (x)\big).
\label{6sc}\eea
Note that the amplitudes in eq. \rf{2sc4grA} and \rf{6sc} have  non-vanishing SSL. We will explain in Sec.~\ref{duality} why all superinvariants in \rf{N8}, \rf{N6}, \rf{N5} have non-vanishing SSL.

\subsection{Details in $\cn=5$, $L=4$ case}

Let us convey here the essential ideas of the derivation of the $\cn=5$ superfield $W_a(x,\theta,\bar\theta)$ from superspace Bianchi identities.  The particle states of the $\cn=8$ theory comprise a single (self-conjugate) multiplet of the superalgebra $OSp(8|4)$. Its helicity states span the range 
$2,3/2,\ldots, -3/2,-2$. For $\cn<8$, there are two conjugate multiplets of $OSp(\cn|4)$, one containing helicities $2,3/2,\dots, 0,\dots -(\cn-4)/2$ and the other the antiparticles. The positive helicity states are described by the (multi-) component spinor fields 
$C_{\a\b\g\d}, \psi^a_{\a\b\g}, F^{ab}_{\a\b}, \l^{abc}_\a,\phi^{abcd},\bar\chi_\da^{abcde}.$
One uses the 5-dimensional Levi-Civita symbol to lower $SU(5)$ indices, e.g.  
$F_{\a\b\,ab} = \e_{abcde}F^{cde}_{\a\b}/3!,$  while two-component spinor indices are raised by $\e^{\a\b}$.

With account of the superspace constraints in \rf{n51}-\rf{n53} one finds that 
there is  a superfield ${\chi}_{\beta}$ which is  anti-chiral and an $SU(5)$ singlet. It was presented  in \cite{Freedman:2017zgq} earlier. In the anti-chiral basis, in which $D^a_\alpha=\partial/\partial\theta^\alpha_a$ and the spacetime coordinate is 
$\bar{y}^{\alpha\dot\alpha}=x^{\alpha\dot\alpha}-\frac12\theta^\alpha_a\bar{\theta}^{\dot\alpha\,a}$, we can write its $\bar\theta$ expansion in the form
\begin{align}
{\chi}^\alpha(\bar{y},\bar{\theta})&=\chi_{678}^\alpha+\bar{\theta}^a_{\dot\alpha}\partial^{\alpha\dot{\alpha}}\bar{\phi}_a+\frac{1}{3!2}\bar\theta^a_{\dot\alpha}\bar\theta^b_{\dot\beta}\epsilon_{abcde}\partial^{\alpha\dot{\alpha}}\bar\lambda^{\dot\beta cde} +\frac1{3!2}\bar\theta^a_{\dot\alpha}\bar{\theta}^b_{\dot\beta}\bar\theta^c_{\dot\gamma} \epsilon _{abcde} \partial^{\alpha\dot{\alpha}}\bar{F}^{\dot\beta\dot\gamma de}\nonumber \\
&+\frac1{4!}\bar\theta^a_{\dot\alpha}\bar\theta^b_{\dot\beta}\bar{\theta}^c_{\dot\gamma}\bar{\theta}^d_{\dot\delta}\epsilon _{abcde} \partial^{\alpha\dot{\alpha}}\bar\psi^{\dot\beta\dot\gamma\dot\delta e}+\frac{1}{5!}\bar{\theta}^a_{\dot\alpha}\bar{\theta}^b_{\dot\beta}\bar{\theta}^c_{\dot\gamma}\bar\theta^d_{\dot\delta}\bar{\theta}^e_{\dot\eta}\epsilon _{abcde} \partial^{\alpha\dot{\alpha}}\bar{C}^{\dot\beta\dot\gamma\dot\delta\dot\eta},
\end{align}
where each component field is a function of $\bar{y}$.  
The first Bianchi identity \rf{n52} tells us that ${W}_a$  has a component expansion of the form
\begin{align}
{W}_a(\bar y,\theta,\bar\theta)=&\phi_a(\bar y)+\theta^\alpha_a{\chi}_{\alpha}(\bar y, \bar\theta)+c_1\bar{\theta}^b_{\dot\alpha}\epsilon_{abcde}\bar\lambda^{\dot\alpha cde}(\bar y)+c_2\bar\theta^b_{\dot\alpha}\bar\theta^c_{\dot\beta}\epsilon_{abcde}\bar{F}^{\dot\alpha\dot\beta de}(\bar y)\nonumber\\
&+c_3\bar\theta^b_{\dot\alpha}\bar\theta^c_{\dot\beta}\bar\theta^d_{\dot\gamma}\epsilon_{abcde}\bar{\psi}^{\dot\alpha\dot\beta\dot\gamma e}(\bar y)+c_4\bar\theta^b_{\dot\alpha}\bar\theta^c_{\dot\beta}\bar\theta^d_{\dot\gamma}\bar\theta^e_{\dot\delta}\epsilon_{abcde}\bar C^{\dot\alpha\dot\beta\dot\gamma\dot\delta}(\bar y).
\end{align}
The value $c_1=1/3!$ is determined by the third Bianchi identity, and the remaining coefficients are fixed using the higher order $ \bar{D}_{a\dot\alpha}$ identities.  Thus we find the ${W}_a$ superfield
\begin{align}
{W}_a(\bar y, \theta,\bar\theta)=&\phi_a(\bar y)+\theta^\alpha_a{\chi}_{\alpha}(\bar y, \bar\theta)+\frac{1}{3!}\bar{\theta}^b_{\dot\alpha}\epsilon_{abcde}\bar\lambda^{\dot\alpha cde}(\bar y)+\frac{1}{2\cdot 2}\bar\theta^b_{\dot\alpha}\bar\theta^c_{\dot\beta}\epsilon_{abcde}\bar{F}^{\dot\alpha\dot\beta de}(\bar y)\nonumber\\
&+\frac{1}{3!}\bar\theta^b_{\dot\alpha}\bar\theta^c_{\dot\beta}\bar\theta^d_{\dot\gamma}\epsilon_{abcde}\bar{\psi}^{\dot\alpha\dot\beta\dot\gamma e}(\bar y)+\frac{1}{4!}\bar\theta^b_{\dot\alpha}\bar\theta^c_{\dot\beta}\bar\theta^d_{\dot\gamma}\bar\theta^e_{\dot\delta}\epsilon_{abcde}\bar C^{\dot\alpha\dot\beta\dot\gamma\dot\delta}(\bar y).\label{Wi}
\end{align}
The superfield $\bar{W}^a$ is the conjugate  and  the 4-loop superspace invariant is then
\bea
S^{L=4}_{\cn=5} =\kappa^{6} \int d^4x d^{10} \theta d^{10}\bar\theta  \,  (  W_a \bar W^a)^3 \,  
\label{N51}\eea
as displayed  schematically in \rf{N5}.

  \section{ Duality Symmetry Action on Scalars and Vectors }\label{duality}
 Full non-linear duality symmetry $\cG$ in $\cN\geq 5$ supergravity acts on scalars and vectors. The number of vectors $A_\mu^{\La}$ with $\La =1, \cdots, n_v$ where $n_v= 10, 16, 28$ for $\cG : SU(1,5), \, SO^*(12), \, E_{7(7)}$ in ${\cal N}=5, 6, 8$, respectively. 
Gravitons are neutral, fermions transform under the compensating $SU(\cN)$ symmetry which has to be combined with duality so that the choice of the local $SU(\cN)$ symmetry is preserved, \cite{Cremmer:1980gs},  \cite{deWit:1981yv}, \cite{Kallosh:2008ic}.

The vector-scalar part of the classical action  is given in the form
$
S=  \int -i F^+ \cN F^+ +c.c .
$
The vector moduli space metric  $\mc N$ depends on scalars $Y$  which are 
inhomogeneous coordinates of the ${\cG\over \cH}$ coset space, for example ${E_{7(7)}\over SU(8)}$ coset space in $\cN=8$ case and
$
\mc N= -i\,\frac{1+Y^{\dag}}{1-Y^{\dag}}$, 
$ Y=B\frac{\tanh \sqrt{B^{\dag}B}}{\sqrt{B^{\dag}B}}$, 
$B_{ij,kl}= -\frac{1}{2\sqrt2}\kappa\,  \varphi_{ijkl}$.
Here $\varphi$ is a canonically normalized field. The $Y$ fields transform as follows
under {\it full non-linear  duality symmetry}, in absence of fermions  \cite{Kallosh:2008ic}
\be\label{deltay}
\delta Y= \Sigma + Y\bar \La - \La Y  - Y  \bar \Si Y.
\ee
The vectors also transform:
\be \label{deltaF}
 \delta F^{+}= [ (\La - \bar \La Y^{\dagger}) + (  \Sigma \, Y^{\dagger} -  \bar \Sigma )]
 \frac{1}{1-Y^\dagger} F^{+}.
\ee
Here $\Lambda$ are parameters of the $\cH$ transformations (e. g.  63 $SU(8)$ transformations in $\cN=8$) and $\Sigma $ are 
 orthogonal  symmetries that extend $\cH$ to $\cG$ (e. g.  70  transformations in $\cN=8$). The $\cG$ group element is
$
E^{-1}=\left(
         \begin{array}{cc}
           1+\Lambda & -\Sigma \\
           -\bar{\Sigma} & 1+\bar{\Lambda} \\
         \end{array}
       \right)
$.
 One would expect that for the asymptotic fields the symmetry following from \rf{deltay}, \rf{deltaF} is 
\be\label{deltay1}
\delta \phi= \Sigma + \phi \bar \La - \La \phi  \qquad 
 \delta F^{+}=  (\La  -  \bar \Sigma ) F^{+}.
\ee
However, one finds that, in fact, the term $ \delta F^{+}=   -  \bar \Sigma  F^{+}$ 
 is inconsistent with the linearized $SU(\cN)$ global symmetry, it beaks it to $SO(\cN)$.
 But since $SU(\cN)$ symmetry is a necessary condition for the asymptotic supersymmetry of the physical states it means that  the $\Sigma$ part of duality symmetry which is orthogonal to $SU(\cN)$ is spontaneously broken.
 The scalars undergo the shift, as well as an $SU(\cN)$ transformations, but the asymptotic vectors do not transform under $\Sigma$ symmetry, only under $SU(\cN)$.
Thus the  immediate consequence of duality on S-matrix is an Adler zero  due to \rf{deltay1}, the part which says that $\delta \phi= \Sigma$. But asymptotic vectors transform only on $SU(\cN)$.

 For $\cN=6,5$ the scalars are a truncated version of $\cN=8$ scalars, and  $\Sigma$ are symmetries associated with the  truncated version of the $E_{7(7)}$ coset. It means that   for 70 $\cN=8$ there are 70 components of $\Sigma$, for 30 scalars of $\cN=6$ there are 30 components of $\Sigma$, 
for 10 scalars of $\cN=5$ there are 10 components of $\Sigma$. 
In all cases, the asymptotic duality symmetry on scalars consists of the linear $SU(\cN)$ and the shift
\be
\delta \phi (x)= \Sigma\, \qquad \delta \bar \phi (x)= \bar \Sigma.
\ee
For linearized superfields describing the asymptotic states of supergravity it means that the superinvariants consistent with asymptotic 
duality have to satisfy an additional symmetry.  Under the constant, $x$ and $\theta, \bar \theta$ independent shifts of the zero dimension superfields the superinvariant has to be invariant under
\be
\delta W(x, \theta, \bar \theta) =  \Sigma.
\ee
Note that all higher components of the on-shell supergravity superfields are contracted with $\theta_i$ and $\bar \theta^i$ and transform only under $SU(\cN)$ symmetry.
This explains why duality symmetry acting on a scalar superfield involves a $\theta,\bar \theta$-independent shift.

This means that only dimension zero superfields are affected by the $\Sigma$-part of duality. All other superfields always have scalars covered by derivatives, and are invariant under asymptotic duality.

Thus our 6-point superinvariants at $L=\cN-1$ have the following symbolic form
\be
\kappa^{2(\cN-2)} \int d^{4\cN}\theta d^4x \, W^6
\label{6pt}\ee
with details in \rf{N8}, \rf{N6}, \rf{N5}. Under duality  we find that the superinvariant transforms, in the symbolic form
\be
 \kappa^{2(\cN-2)}  \int d^{4\cN}\theta d^4x \,  \delta W^6
\sim  \kappa^{2(\cN-2)}  \Sigma \int d^{4\cN}\theta d^4x \,  W^5
\label{d6pt}\ee
and it means that it breaks duality, so the SSL is non-vanishing.
In more detail we find that 
\be
\kappa^{2(\cN-2)} \delta \int d^{4\cN}\, \theta d^4 x \,   \Tr \Big ((W \bar W)^3 \Big )  =  3  \kappa^{2(\cN-2)}  \int d^{4\cN}\, \theta d^4 x \,   \Tr \Big ( (\Sigma \bar W + W\bar \Sigma) (W \bar W)^2 \Big ),
\label{D4R6delta}\ee
\be
\kappa^{2(\cN-2)} \delta \int d^{4\cN}\, \theta d^4 x  ( \Tr W \bar W)^3=
3 \kappa^{2(\cN-2)}  \int d^{4\cN}\, \theta d^4 x \,     \Big ( \Tr (\Sigma \bar W + W\bar \Sigma) (\Tr W\bar W)^2 \Big ) . \label{D4R62}\ee
The superspace integral of 5 superfields does not vanish
\be
 \kappa^{2(\cN-2)}  \Sigma \int d^{4\cN}\theta d^4x \,  W^5 (x, \theta,\bar \theta)\neq 0.
\label{d6pt1}\ee
This is in the contrast with 4-point amplitudes in \rf{3loop}, \rf{HST} where 
\be
\delta (CT_{L=3}^{\cN}) =\kappa^4 \Sigma \int d^4 x \, d^{2\cN} \theta \,  W^3 (x, \theta,\bar \theta) =0.
\ee
since for all massless fields any 3-point vertex vanishes by the momentum conservation.

Thus, all our 6-point superinvariants in \rf{N8},  \rf{N6},  \rf{N5} break asymptotic duality. Namely,  amplitudes containing scalars have a non-vanishing SSL.

\section{Superamplitudes for $\cn=5,6,8$ supergravity}\label{N5SA}

The superamplitude approach to $\cn=8$ is well known \cite{Elvang:2010jv,Beisert:2010jx}. An $n$-particle  ${\text N}^k\text{MHV}$  superamplitude is a generating function from which all amplitudes related by supersymmetry in  a given $n$-particle  ${\text N}^k\text{MHV}$ family can be obtained by differentiation with respect to Grassmann valued bookkeeping variables $\eta_{ia},\, i=1,\dots,n  , \,a=1,\dots,8.$  A succinct recipe was given in \cite{Elvang:2011fx} to obtain superamplitudes for $\cn < 8$ from those for $\cn=8$ by supressing $\eta_{ia}$ variables with SU($\cn$) indices in the range $a=\cn+1,\cn+2,\dots,8$.  Alternatively, one can deal with directly with $\cn <8$ and apply the method of Secs. 3.1 and 3.3 of \cite{Elvang:2009wd} to construct the superamplitude as an expansion in terms of basis amplitude for component particles.

Superamplitudes\footnote{We use the conventions of the book \cite{Elvang:2015rqa}.}  are constructed from 2-component spinors $|i\>,\,|i]$ that encode placement within the amplitude and their momenta $p_i=- |i\>[i|-|i]\<i|$.  The angle and square spinors are the momentum space wave functions of free Weyl fermions and therefore have mass dimension 1/2.   Amplitudes and superamplitudes that correspond to  candidate counterterms must be \emph{local} functions, that is polynomials, in the Lorentz invariant brackets $\<ij\>,\,[kl]$. Locality plays a basic role in determining the structure and consistency of superampitudes.  Finally we note that ${\text N}^k\text{MHV}$ superamplitudes in $\cn$-extended supergravity are  polynomials of order $(k+2)\cn$ in the bookkeeping variables $\eta_{ia}$.  Component amplitudes are projected by applying an order $(k+2)\cn$ derivative Grassmann derivative (as we discuss for $\cn=5$ supergravity in Sec \ref{N5NMHV} below.)

\subsection{$4$-point MHV superamplitudes}
MHV superamplitudes are the simplest; they require the basic SUSY invariant 
\begin{align} 
\delta^{(\cn)}(\tilde{Q})= &\frac{1}{2^\cn}\prod_{a=1}^\cn\sum_{i=1}^n \<ij\>\eta_{ia}\eta_{ja},
\label{mhvnpt}
\end{align}
and take the form\footnote{The factor  $\d^{(4)}(\sum_i p_i)$ which expresses momentum conservation is omitted in all superamplitudes discussed in this paper.} 
\be
C_{MHV}^{\cn,n}  = \frac{c_n}{\<12\>^\cn} \delta^{(2\cn)}(\tilde{Q}).
\label{gramp}
\ee
The constant $c_n$ is the $n$-graviton amplitude, usually denoted by $\<--++\dots +\>$ in the ordering convention that states of the -ve helicity multiplet must appear in positions 1 and 2.  To see this, one simply
applies the appropriate Grassmann derivative:
\be
\prod_{a=1}^5 \frac{\pa^2}{\pa_{\eta_{1a}}\pa_{\eta_{2a}}}C_{MHV,n} = c_n \equiv \<--++\dots +\>\,.
\ee
Following \cite{Bianchi:2008pu,Kallosh:2008ru}, the 4-point $L=3$ case, commonly called the $R^4$ superamplitude, can be written for all $\cN\geq 5$ as
\bea
C_{MHV}^{\cn,4} &=& \delta^{(\cn)}(\tilde{Q})\frac{[34]^4}{\<12\>^{\cN-4}},\\
\<--++\>&=& \<12\>^4[34]^4.
\eea
At loop level $L=4$ we are interested in $\pa^2R^4$, and an important difference between $\cn=8$ and $\cn<8$ emerges.  The $\cn=8$ theory contains a single CPT self-conjugate multiplet, so derivatives are taken symmetrically on all 4 lines. The superamplitude and graviton amplitude then acquire the factor $s+t+u$ and thus vanish on-shell.   
For $\cn<8$ the -ve and +ve helicity states are in different supermultiplets,  and one has separate permutation symmetry under $1\lra2$ and $3\lra4$ exchanges\footnote{ For $n>4$, invariance under permutations of the set $3,4\dots,n-1,n$ is required.}.   The superamplitude is then
\bea
C_{MHV}^{\cn,4} &=& \delta^{(\cn)}(\tilde{Q})s \frac{[34]^4}{\<12\>^{\cN-4}},\\
\<--++\>&=&s \<12\>^4[34]^4.
\eea

\subsection{NMHV 6-point amplitudes for $\cn=5$ SG}\label{N5NMHV}
As explained  at the beginning of Sec.~\ref{NMHVsix},  there are 5-point local super-invariants at loop level $L=\cN$ and beyond,  and the same is true for superamplitudes.  In both cases their SSL's vanish. We therefore proceed to the study of 6-point superamplitudes in this section.  We will derive an interesting $L=4$ superamplitude with non-vanishing SSL.

We  start by defining super wavefunctions $\Phi$ and $\Psi$ for the +ve and -ve helicity multiplets of the theory\footnote{Note that the +ve (and -ve) helicity multiplets include singlet fermions $\chi_-$ (and $\bar\chi_+$) with reverse helicity.}: 
\begin{align}
\Phi=&h_++\eta_a\psi^a_++\frac{1}{2!}\eta_a\eta_bv_+^{ab}+\frac{1}{3!}\eta_a\eta_b\eta_c\lambda^{abc}_++\frac{1}{4!}\eta_a\eta_b\eta_c\eta_d\epsilon^{abcde}\phi_e+\frac{1}{5!}\eta_a\eta_b\eta_c\eta_d\eta_e\epsilon^{abcde}\chi_-,\\
\Psi=&\bar{\chi}_++\eta_a\bar{\phi}^a+\frac{1}{2!3!}\eta_a\eta_b\epsilon^{abcde}\lambda_{cde-}+\frac{1}{3!2!}\eta_{a}\eta_b\eta_c\epsilon^{abcde}\bar{v}_{de-}+\frac{1}{4!}\eta_a\eta_b\eta_c\eta_d\epsilon^{abcde}\bar{\psi}_{e-}\nonumber\\
&+\frac{1}{5!}\eta_a\eta_b\eta_c\eta_d\eta_e\epsilon^{abcde}h_-.
\end{align}
The $\eta$ variables in the wave functions determine the derivatives used to project out the corresponding particle state in a component amplitude. For a +ve helicity gravitino state, one uses
$\pa/\pa_{\eta_a}$; for a -ve helicity photino $\l^{ab}_-$ one needs $\pa^2/\pa_{\eta_a}\pa_{\eta_b}$, etc.  Positive helicity gravitons and singlet photinos require no derivatives.

NMHV superamplitudes depend on the invariants $\d^{(\cn)}(\tQ)$ and products of 
\be
m_{ijk,a} = [ij]\eta_{ka}+[jk]\eta_{ia}+[ki]\eta_{ia}.\label{basic}
\ee
The general  6-point  $\cn=5$ NMHV superamplitude is a superposition of 6 products of the basic invariants.  This basis expansion was first obtained for $\cn=8$ supergravity in \cite{Elvang:2009wd}. In  Appendix \ref{NMHVgeneral}, we adapt the method to the $\cn=5$ theory and
 start here with the result of that construction:
\begin{align}
{\cal M}_6=& c_1 X_{11111}+c_2 X_{(11112)}+c_3 X_{(11122)}\nonumber\\
&+c_4 X_{(11222)}+c_5 X_{(12222)}+c_6 X_{22222},\label{m61}
\end{align}
where $X_{ijklm}$ is given by
\begin{equation}
X_{ijklm}=\frac{\delta^{(10)}(\tilde{Q})m_{i34,1}m_{j34,2}m_{k34,3}m_{l34,4}m_{m34,5}}{[34]^5\langle56\rangle^5}.\label{defX}
\end{equation}
The (..) in \reef{m61} indicate symmetrized ordering of the indices 1,2, e.g $X_{(11112)}$ contains 5 terms and $X_{(11122)}$ contains 10.  The coefficients $c_i$ are each 6-point amplitudes in the ordering convention (used in the $\cn=8$ literature) that negative helicity states are placed at positions 1,5,6.
To see this just apply the appropriate derivatives to obtain
\begin{align}
{\cal M}_6=& \langle-+++--\rangle X_{11111}+\langle\psi_-^{1234}\psi_-^{5}++--\rangle X_{(11112)}+\langle v_-^{123}v_+^{45}++--\rangle X_{(11122)}\nonumber\\
&+\langle\lambda_-^{12}\lambda_+^{345}++--\rangle X_{(11222)}+\langle\bar{\phi}^1\phi^{2345}++--\rangle X_{(12222)}+\langle\chi_+\chi_-++--\rangle X_{22222}.\label{genbas}
\end{align} 
To summarize, the 6-point NMHV superamplitude is determined by only  six basis matrix elements. This construction is valid for $\cn=5$ supergravity, independent of loop order.

\subsection{$\mathcal N =5 \ L=4$ six-point superamplitude}\label{N5L4six}
The only properties used to construct  this basis are $\cn=5$ supersymmetry and its Ward identities plus $SU(5)$ R-symmetry. The basis amplitudes might describe any realization of these properties. For example, they might describe the 6-point NMHV tree level amplitudes calculated using BCFW recursion relations; in this case they would have non-local pole terms.  In our application to candidate counterterms, they must be \emph{local}, that is polynomials in the $\<ij\>,\,[kl]$  brackets of total mass dimension fixed by the  particular loop order under study.  Locality, dimensions, helicity weights, and identical particle symmetries provide very strong contraints on these polynomials. They tell us that certain basis amplitudes vanish and the others are determined up to a small number of constant parameters. In the case $L=4$ the total dimension is 10, so the total number of  angle and square spinors is $\sum_i (a_i+s_i) = 20.$  The helicity weight constraint requires that for each of the $6$ particles the difference $(a_i-s_i) = -2h_i$, where $h_i$ is the helicity. The bose symmetry constraint requires that the basis amplitudes are invariant under the exchange $3\leftrightarrow4$ of spinors (and momenta) and the same for $5\leftrightarrow6.$ 

Let's apply these constraints to the basis matrix elements in \reef{genbas}.  For $ \langle-+++--\rangle$, the helicity weight tally is $|1\>^4,\,|2]^4,\,|3]^4,\,|4]^4,\,|5\>^4,\,|6\>^4$.  Thus the minimal number of spinors needed to describe the 6-graviton amplitude is 24. The total mass dimension, namely 12, exceeds $\D=10$, so this matrix element vanishes.  A similar argument shows that the gravitino basis amplitude $\langle\psi_-^{1234}\psi_-^{5}++--\rangle$ also vanishes.  The vector amplitude is more subtle. Its spinor count is
$|1\>^2,\,|2]^2,\,|3]^4,\,|4]^4,\,|5\>^4,\,|6\>^4$ which is allowed by the $\D=10$ constraint.  However, the unique coupling of these spinors is
\begin{equation}
\langle v_-v_+++--\rangle=c_1\langle 15\rangle \langle 16\rangle\langle 56\rangle^3[23][24][34]^3,\label{vecamp}
\end{equation} 
and this is \emph{odd} rather than \emph{even} under the exchanges $3\leftrightarrow4$ and $5\leftrightarrow6$.  So the vector basis element must also vanish.

Next is the $\l$ basis amplitude. Its helicity weights sum to dimension 9, so one has to include a momentum $p$ in the ansatz for the matrix element: 
\be
\langle\lambda_-\lambda_+ ++--\rangle = \<1| p |2] [34]^4\<56\>^4, \qquad p =c_2(p_3+p_4).
\ee
This form is unique, since other possibilities can be eliminated using the Schouten identity.  The momentum must be invariant under the exchanges.  Using momentum conservation, we can adopt $p=c_2(p_3+p_4)$ where $c_2$ is a constant.  The same argument applies to the singlet $\chi$ basis element, so we can write
\be
\langle\chi_+\chi_-++--\rangle = [1|q|2\>[34]^4\<56\>^4,\qquad q=b(p_3+p_4)\,.
\ee
Finally we must specify the scalar basis element.  After elimination of other forms using Schouten, we arrive at the ansatz
\be
\langle\bar{\phi}^1\phi^{2345}++--\rangle = s_1 [34]^4\<56\>^4,\label{phibas}
\ee
where $s_1$ is a quadratic invariant which must be even under exchanges.  At this point there are several possibilities, such as $s_1=c p_1\cdot(p_5+p_6)+c'p_3\cdot p_4$, so we keep an open mind.  

This concludes the first phase of the analysis.  We have found local expressions  for the surviving 3 basis amplitudes consistent with the constraints discussed above. So the basis expansion reduces to
\begin{equation}
{\cal M}_6=\langle\lambda_-\lambda_+++--\rangle X_{(11222)}+\langle\bar{\phi}\phi++--\rangle X_{(12222)}+\langle\chi_+\chi_-++--\rangle X_{22222}.\label{basis3}
\end{equation}
 
For full consistency \emph{all} amplitudes obtained by applying  independent 15th order products of $\eta_{ia}$ derivatives to 
\reef{basis3} must be local.  To begin this process we study permuted \emph{basis elements}. The allowed permutations of position must respect the ordering convention that positve helicity particles appear at positions 2,3,4.  The  permuted basis element obtained from the basis expansion must be local and reproduce the permutation of the original form.
Let's examine the permutation 
\be
\langle\lambda^{12}_-++\lambda_+^{345}--\rangle=\partial_{11}\partial_{12}\partial_{43}\cdots\partial_{45}\partial_{51}\cdots\partial_{55}\partial_{61}\cdots\partial_{65}\mathcal{M}_6.
\ee
Only  the unpermuted $\l$ basis element has non-vanishing 15th derivative leaving
\begin{align}
\langle\lambda^{12}_-++\lambda_+^{345}--\>
=&\frac{[23]^3}{[34]^3}\langle\lambda^{12}_-\lambda_+^{345}++--\rangle\nonumber\\
=&-c_2\langle1|(p_5+p_6)|2][34][23]^3\langle56\rangle^4. \label{lamperm}
\end{align}
In the last step, we used momentum conservation. 
The result is local, but we need more; the permutation $2\leftrightarrow4$ should give the amplitude
\begin{equation}
\langle\lambda^{12}_-++\lambda_+^{345}--\rangle=-c_2\langle1|p_5+p_6|4][23]^4\langle56\rangle^4.\label{perm24}
\end{equation}
The difference  between \reef{perm24} and \reef{lamperm} should vanish, so we form the difference, extract common factors and find
\bea
&&-c_2\langle1|p_5+p_6|4][23]^4\langle56\rangle^4+c_2\langle1|(p_5+p_6)|2][34][23]^3\langle56\rangle^4\nonumber\\
&&\qquad=-c_2[23]^3\<56\>^4\<1|(5+6)\bigg(|4][23]-|2][34]\bigg),\label{perm}
\eea
which vanishes only if $c_2=0$.   So the basis reduces to two terms! 

To consider permuted scalar matrix elements, we use the (temporary) notation  
$A_{1k}=a_{1k}[34]^4\<56\>^4$ for the three cases with $\bar\phi^5$ at position 1 and $\phi^{2345}$ at positions $k=2,3,4$.  Applying the appropriate derivatives to ${\cal M}_6$ we learn that only the scalar channel contributes, and that $a_{12}=a_{13}=a_{14}$.  Similarly, we write $A_{j2} = a_{j2}[34]^4\<56\>^4$ for the cases where  $\phi^{2345}$  is at position 2 and $\bar\phi^5$ is at positions  $j=1,5,6$. We then derive  $a_{12}=a_{52}=a_{62}$.  Thus the quadratic invariant $s$ in \reef{phibas} must be invariant under permutations of $p_2,p_3,p_4$ and  $p_1,p_5,p_6$. Thus, the locality and consistent permutation properties lead to $s =-a(p_1+p_5+p_6)\cdot(p_2+p_3+p_4) = a(p_1+p_5+p_6)^2=a(p_2+p_3+p_4)^2$.
The double permutation $\<-++\phi^{2345}\bar\phi^1-\>$ is also consistent.

Next, we consider the singlet fermion permutation. The permuted amplitude can be calculated as
\begin{align}
\langle\chi_+++\chi_---\rangle=&\partial_{41}\cdots\partial_{45}\partial_{51}\cdots\partial_{55}\partial_{61}\cdots\partial_{65}{\mathcal M}_6\nonumber\\
=&\frac{\langle56\rangle^4[23]^4}{[34]}(5s[13]+[23][1|q|2\rangle).
\end{align}
Note that the factor 5 comes from $X_{(12222)}$. Let us take $q=b(p_3+p_4)$ and $s=a(p_2+p_3+p_4)^2$. Note that since there is $1/[34]$ pole, we must eliminate it by choosing $c_3$ and $c_4$ properly. The pole can be removed if we assume $5a=-b$. Then, we find that
\begin{equation}
5[13]s+[23][1|q|2\rangle=-5a[34][1|(p_2+p_3)|4\rangle.
\end{equation}
Thus, we obtain
\begin{equation}  
\langle\chi_+++\chi_---\rangle=-5a[1|(p_2+p_3)|4\rangle[23]^4\langle56\rangle^4.\label{chiperm1}
\end{equation}
This is a consistent permuted amplitude.  We also checked the consistency of the $1\lra6$ permutation of the $\chi$ basis element.  After spinor algebra needed to show that the initial $1/\<56\>$ pole cancels we find
\begin{align}
 \langle -\chi_-++-\chi_+\rangle
 =&5a[6|(p_3+p_4)|2\rangle[34]^4\langle15\rangle^4.\label{perm2}
\end{align}
Comparison with \reef{chiperm1} reveals a small subtlety. The two forms are properly related by the permutation $1\leftrightarrow6$ including the - sign from fermion exchange.

We summarize the result:
\begin{equation}
{\cal M}_6=\langle\bar{\phi}\phi++--\rangle X_{(12222)}+\langle\chi_+\chi_-++--\rangle X_{22222},\label{N5L4sa}
\end{equation}
where
\begin{equation}
\langle\bar\phi\phi++--\rangle=a(p_1+p_5+p_6)^2[34]^4\langle56\rangle^4,\label{scbas}
\end{equation}
\begin{equation}
\langle\chi_+\chi_-++--\rangle=-5a [1|(p_3+p_4)|2\rangle[34]^4\langle56\rangle^4,\label{chibas}
\end{equation}
and 
\begin{equation}
X_{ijklm}=\frac{\delta^{(10)}(\tilde{Q})m_{i34,1}m_{j34,2}m_{k34,3}m_{l34,4}m_{m34,5}}{[34]^5\langle56\rangle^5},
\end{equation}
where $m_{i34,a}=[i3]\eta_{4a}+[34]\eta_{ia}+[4i]\eta_{3a}$.  We believe that this superamplitude corresponds to the 6-point superspace invariant \reef{N51}.  The superamplitude and superspace formalisms look rather different, but they are both based on the same principles of linearized $\cn=5$ supersymmetry.  The advantage of superspace is compactness; the advantage of superamplitudes is the explicit algorithm (via 15th order $\eta$-derivatives) provided for the computation of component S-matrix elements.  We proceed under the assumption that they carry the same physics.

The main motivation for our work is the question of the soft limit of S-matrix elements containing scalar particles obtained from \reef{N5}. The corresponding S-matrix elements shown in  eqs. \rf{2sc4grA}, \rf{6sc} have  non-vanishing SSL. The general reason for this was explained in Sec. 4, as a property of the dimensionless superfields to shift under asymptotic duality.

 Here  we can consider the same question for amplitudes projected from 
\reef{N5L4sa}.  We examine the scalar basis element \reef{scbas} together with two amplitudes:
\begin{align}
\langle\bar\phi\phi++--\rangle=&a(p_1+p_5+p_6)^2[34]^4\langle56\rangle^4,\label{scbas}\\
\langle\bar\phi\phi\phi\phi\bar\phi\bar\phi\rangle=&  as^5,\\
\langle \chi_++\psi^1_+\phi^{2345}--\rangle
=&a\langle56\rangle^4[23]^3(s[12]-5[24][1|(p_2+p_3)|4\>),
\end{align}
with $s = (p_2+p_3+p_4)^2$.  One sees immediately that their single soft limits do not vanish. The last result shows  that gravitinos are present in the spectrum, although absent in the basis \reef{N5L4sa}.  The same feature for vectors can be seen from
\begin{align}
\langle \chi_++\lambda^{123}_+v_+^{45}--\rangle=a\langle56\rangle^4[24]^2[23](3s[12]-5[24][1|(p_2+p_3)|4\>).
\end{align}

In the last two examples, the initial expression obtained from their Grassmann derivatives applied to \reef{N5L4sa} contained a $1/[34]$ pole, and spinor manipulations using Schouten and momentum conservation were used to show that it cancels.  

\subsection{All 6-point $L=4$ $\cN=5$ amplitudes are local}\label{locality}
As we stated below \reef{basis3}, complete consistency requires that \emph{all} amplitudes computed from \reef{N5L4sa} are local.  Explicit computation of all amplitudes is an impractical task\footnote{A rough estimate of the total number of independent amplitudes is 200; there were already 51 amplitudes in the simpler $\cn=8$, $L=3$ $R^4$ invariant \cite{Freedman:2011uc}.},  so a general proof of locality, independent of explicit amplitudes, is desirable.  In this section we provide a proof. The key idea (see Sec. 2.3 of \cite{Elvang:2010jv}) is that any superamplitude whose  basis  amplitudes are local and behave consistently under permutation produces local matrix elements for all processes if it has full permutation symmetry.  For the 6-point NMHV superamplitude \reef{genbas}, full permutation symmetry means that the superamplitude is invariant under the separate $S(3)$ groups of permutations of position of its +ve and -ve helicity states, i.e for $234$ and $156$.

To explain this further, we rewrite \reef{N5L4sa} as
\begin{align}
{\cal M}_6=& \frac{a\d^{10}(\tQ)}{[34]\<56\>}\bigg[(p_2+p_3+p_4)^2 Y_{(12222)} - 5[1|p_3+p_4|2\>Y_{22222}\bigg]\\
Y_{ijklm}&= m_{i34,1}m_{j34,2}m_{k34,3}m_{l34,4}m_{m34,5}\\
m_{i34,a}=&[i3]\eta_{4a}+[34]\eta_{ia}+[4i]\eta_{3a}.
\label{3456bas}
\end{align}
\noindent By inspection we see two things:\\
i.  There is evident symmetry under $3\lra4$ and $5\lra6$, but full symmetry in 234 and 156 is not manifest,\\
ii.  The only possible singularities are first order poles $1/[34]$ and $1/\<56\>$.\\
\noindent The point is that these singularities are artificial and have essentially no relation to the physics. With reference to the discussion of the basis expansion in Appendix~\ref{NMHVgeneral},  we see that those factors arose because two lines, namely 5, 6, were chosen to exploit $\tQ_a$ supersymmetry and two more, namely 3, 4, were needed to enforce $Q^a$ supersymmetry.  

A  physically equivalent  representation of the superamplitude, let's call it $\tilde{{\cal M}}_6$,   can be derived by choosing any distinct pair of lines $q,r$ for $\tQ_a$ SUSY and a different pair $s,t$ for $Q^a$ SUSY.   Of course we would need permuted basis amplitudes (and $Y$-polynomials) appropriate to the new choices. Then matrix elements derived from $\tilde{{\cal M}}_6$ have possible singular factors $1/[st]$ and $1/\<qr\>$ rather than $1/[34]$ or $1/\<56\>$.

Suppose that we could prove  that ${\cal M}_6=\tilde{{\cal M}}_6$, by which we mean that all matrix elements obtained from the first agree with those obtained from the second. Then we would know that the singular factors $1/[34]$ and $1/\<56\>$ are absent in all physical amplitudes.  
Actually, since the basis elements uniquely determine the full superamplitude, it is sufficient to check that the permuted basis amplitudes of $\tilde{{\cal M}}_6$ are produced by acting on $\mathcal{M}_6$ with the Grassmann differential operators for the permutation.  However, this is something we already did when  we verified the consistency of permuted  basis elements in the previous section.  There we worked out all perturbations of the scalar basis element and the $2\lra4$ and $1\lra6$ exchanges of the singlet fermion element. Together with the manifest $34$ and $56$
symmetries of \reef{3456bas}, this gives sufficient information to conclude that full permutation symmetry obtains.

\subsection{No $L=3,$ $\cN=5$ local six-point amplitude}
In this section we show that a  local $L=3$ analogue of the previous 6-point NMHV invariant does not exist.
We can use the general formula~\eqref{genbas}, noting that each basis amplitude must be local and have dimension 8. Then,  simply from dimension counting,  
one   finds immediately that the only basis element permitted  is 
\begin{equation}
\langle\bar\phi\phi++--\rangle=a[34]^4\langle56\rangle^4.
\end{equation}
The six-point superamplitude  then has the single term
\begin{equation}
\mathcal{M}_6=a\,[34]^4\langle56\rangle^4X_{(12222)}.\label{L3supamp}
\end{equation}

However, it is easy to show that this form is inconsistent.  
Dimensional analysis does not allow a local form for  $\langle \chi_+\chi_-++--\rangle$, so this basis amplitude vanishes.  For consistency we must require that  permutations also vanish.
So we consider the permuted amplitude $\langle \chi_+++\chi_---\rangle$ and compute it by applying the appropriate $\eta$ derivative to the candidate superamplitude \reef{L3supamp}:
\begin{equation}
\langle \chi_+++\chi_---\rangle=\partial_{41}\cdots\partial_{46}\partial_{51}\cdots\partial_{56}\partial_{61}\cdots\partial_{66}\mathcal M_6=\frac{a[13][23]^4\langle56\rangle^4}{[34]}.
\end{equation}
Consistency requires $$a=0.$$  One can also show that the amplitude $\langle\chi_++\lambda^{123}_+v_+^{45}--\rangle$ is non-local unless $a=0$.
The conclusion is that there is no  6-point local counter term for $\mathcal N=5,\ L=3$.

\subsection{$\cn=6$ NMHV superamplitude}\label{N6SA}

States of +ve helicity multiplet of  $\mathcal N =6$ supergravity span the range between the helicity 2 graviton and a helicity -1 singlet vector. These states and their -ve helicity counterparts are joined in the following super wave functions: 
\begin{align}
\Phi=&h_++\eta_a\psi^a_++\frac{1}{2!}\eta_a\eta_bv_+^{ab}+\frac{1}{3!}\eta_a\eta_b\eta_c\lambda^{abc}_++\frac{1}{2!4!}\eta_a\eta_b\eta_c\eta_d\epsilon^{abcdef}\phi_{ef}\nonumber\\
&+\frac{1}{5!}\eta_a\eta_b\eta_c\eta_d\eta_e\epsilon^{abcdef}\chi_{f-}+\frac{1}{6!}\eta_a\eta_b\eta_c\eta_d\eta_e\eta_f \bar{v}_{78-},\\
\Psi=&v^{78}_++\eta_a\bar{\chi}_+^a+\frac12\eta_a\eta_b\bar{\phi}^{ab}+\frac{1}{3!3!}\eta_a\eta_b\eta_c\epsilon^{abcdef}\bar\lambda_{def-}+\frac{1}{4!2!}\eta_{a}\eta_b\eta_c\eta_d\epsilon^{abcdef}\bar{v}_{ef-}\nonumber\\
&+\frac{1}{5!}\eta_a\eta_b\eta_c\eta_d\eta_e\epsilon^{abcdef}\bar{\psi}_{f-}+\frac{1}{6!}\eta_a\eta_b\eta_c\eta_d\eta_e\eta_f\epsilon^{abcdef}h_-.
\end{align}

We consider the six-point NMHV amplitude, which is formally written as $\langle \Psi\Phi\Phi\Phi\Psi\Psi\rangle$. The starting point of our analysis is the general basis expansion (which is similar but not identical to \reef{genbas}): 
\begin{align}
{\cal M}_n=& \langle-+++--\rangle X_{111111}+\langle\psi_-^{12345}\psi^{6}_+++--\rangle X_{(111112)}+\langle v_-^{1234}v_+^{56}++--\rangle X_{(111122)}\nonumber\\
&+\langle\lambda_-^{123}\lambda_+^{456}++--\rangle X_{(111222)}+\langle\bar{\phi}^{12}\phi^{3456}++--\rangle X_{(112222)}\nonumber\\
&+\langle\chi_+^1\chi_-^{23456}++--\rangle X_{(122222)}+\langle{v}^{78}_+\bar{v}_{78-}++--\rangle X_{222222},
\end{align}
where
\begin{equation}
X_{ijklmn}=\frac{\delta^{(12)}(\tilde{Q})m_{i34,1}m_{j34,2}m_{k34,3}m_{l34,4}m_{m34,5}m_{n34,6}}{[34]^6\langle56\rangle^6},
\end{equation}
and $m_{i34,a}=[i3]\eta_{4a}+[34]\eta_{ia}+[4i]\eta_{3a}$.

\subsection{No $L=3,\,4$ local six-point superamplitudes}
The mass dimension of the basis amplitudes is $2(L+1)=10 $ (independent of $\cn$), so we can take over the results from the $\cn=5$ section that there are only three 
possible nonzero basis amplitudes. These take the form:
\begin{equation}
\langle\lambda_-^{123}\lambda_+^{456}++--\rangle=a\langle1|(p_3+p_4)|2][34]^4\langle 56\rangle^4,
\end{equation}
\begin{equation}
\langle\bar{\phi}^{12}\phi^{3456}++--\rangle=s[34]^4\langle 56\rangle^4,
\end{equation}
\begin{equation}
\langle\chi_+^1\chi_-^{23456}++--\rangle= b[1|(p_3+p_4)|2\rangle[34]^4\langle 56\rangle^4,
\end{equation}
where $a$ and $b$ are undetermined constants and $S$ is a momentum square that is symmetric under $3\leftrightarrow4$ and $5\leftrightarrow6$.
The discussion of the permuted $\l$ and scalar basis amplitudes is the same as for $\cn=5$, so again we find that
$a=0$ and $s=c(p_1+p_5+p_6)^2$.

Next we consider the permuted  $\chi$ basis element,
\begin{align}
\langle \chi^1_+++\chi_-^{23456}--\rangle=&\partial_{11}\partial_{42}\cdots\partial_{46}\partial_{51}\cdots\partial_{56}\partial_{61}\cdots\partial_{66}\mathcal{M}_6\nonumber\\
=&\frac{[23]^4\langle56\rangle^4}{[34]}(b[1|(p_3+p_4)|2\rangle[23]+5S[13]).
\end{align}
The factor $5$ comes from symmetrization in $X_{(112222)}$. As in the $\mathcal{N}=5$ case, we can remove the $1/[34]$-pole by choosing
\begin{equation}
-b=5c.
\end{equation} 
Then, we find
\begin{equation}
\langle \chi^1_+++\chi_-^{23456}--\rangle=b[1|p_2+p_3|2\rangle[23]^4\langle56\rangle^4.
\end{equation}
This has the correct permuted structure, so we seem to find a local $\mathcal N=6,\ L=4$ six-point counterterm.

However, there is one more important check to make. As in $\cn=5$,  locality, permutation symmetry, and little group scaling require that the amplitude $\langle v^{78}_+++\bar{v}_{78-}--\rangle$ vanishes.  Consistency requires that its permutations also vanish. So we consider
\begin{align}
\langle v^{78}_+++\bar{v}_{78-}--\rangle=&\partial_{41}\cdots\partial_{46}\partial_{51}\cdots\partial_{56}\partial_{61}\cdots\partial_{66}\mathcal M_6\nonumber\\
=&\frac{15c[13][23]^4\langle56\rangle^4}{[34]^2}\left((p_2-p_3-p_4)^2[13]+[14]\langle42\rangle[23]\right).
\end{align}
This amplitude vanishes if and only if $c=0$. Therefore, all  amplitudes vanish!

 A similar argument rules out $L=3$ six-point superamplitude. Therefore, we conclude that there is no local six-point superamplitudes in $\mathcal N=6$ supergravity at loop levels $L=3, 4$. There is nothing to cancel the expected non-vanishing soft limits of the supersymmetric completion of the 4-point MHV candidate counterterms, so these candidates are ruled out and we predict ultraviolet finiteness.


 \section{Summary}\label{summary}

The main new ingredients in our work were 6-point full superspace invariants constructed from constrained dimensionless superfields $W(x,\theta,\bar\theta)$ and the corresponding superamplitudes. These invariants can be constructed beginning at loop level $L =\cn-1$. The scalar amplitudes obtained from them have non-vanishing soft scalar limits.   The $\cn=5,6,8$ supergravities all have 4-point invariants whose nonlinear completions also have 6-point amplitudes with non-vanishing SSL's. The non-vanishing property was established for $\cn=8$ supergravity by \cite{Elvang:2010kc,Beisert:2010jx}, and it is reasonable to assume that it is present for $\cn=5,6$ as well.  The scalar fields of $\cn=5,6,8$ supergravities are Goldstone bosons of ${\cal G}/{\cal H}$ cosets, so that vanishing SSL's are required.

A UV divergence in a 4-point amplitude would require the addition of a 4-point invariant to the Lagrangian, and the ensuing non-vanishing SSL's must be cancelled.
For $L<\cn-1$ a UV divergence in 4-point amplitudes can be ruled out because there is no 6-point invariant available to cancel the SSL's. But at loop order $L=\cn-1$, cancellation may occur and we cannot draw a firm conclusion from the SSL argument.

With this strategy we confirm in $\cn=8$ supergravity the results of \cite{Elvang:2010kc, Beisert:2010jx}  that the SSL argument based on \E\, symmetry implies UV finiteness for $L<7$ but not beyond. For the $\cN=6$ model, our results show that the SSL argument to $SO^*(12)$ symmetry requires UV finiteness for $L=3,4$, but not for $L=5$.  UV finiteness at $L=3,4$ is thus a prediction. For $\cN=5$ model we  found that   the single soft scalar limit due to $SU(1,5)$ symmetry does not necessarily require  the $L= 4$  UV finiteness, but does require it for $L<4$. Specifically for $L=3$, this provides the first known explanation of the 3-loop part of the computation in \cite{Bern:2014sna}. But since $\cN=5$, $L= \cN-1=4$ was also found to be  UV finite in \cite{Bern:2014sna}, a new study is required to explain this computation and the cancellation of the corresponding 82 diagrams.

Our results follow from two independent investigations,  one via superspace invariants, the other by superamplitude methods. It is gratifying that all information obtained from these independent techniques agrees perfectly.

\
 
 \noindent{\bf {Acknowledgments:}} We are grateful to  Z. Bern, L. Dixon,  and  especially to H. Elvang  for  useful  discussions of the current  work  and to H. Nicolai and R. Roiban for a collaboration on a related project.  This work  is supported by SITP and by the US National Science Foundation grant PHY-1720397.  The work of DZF is partially supported by US NSF grant Phy-1620045. 

\appendix

\section{NMHV basis expansion for ${\mathcal N}=5$ superamplitudes}\label{NMHVgeneral}

This construction follows  \cite{Elvang:2009wd} closely.
The  NMHV superamplitude is a 15th order polynomial in its bookkeeping variables $\eta_{ia}$, so we start with the generic form
\begin{equation}\label{N5gen}
\mathcal{M}^{NMHV}_n=P_5\delta^{(10)}(\tilde{Q}_a),
\end{equation}
with
\begin{equation}
P_5=\sum_{i,j,k,l,m=1}^nq_{ijklm}\eta_{i1}\eta_{j2}\eta_{k3}\eta_{l4}\eta_{m5},
\end{equation}
where $q_{ijklm}$ is a function of momentum spinors. $P_5$ is manifestly invariant under $\tilde{Q}$, and we will require $Q$-invariance later to determine the $q_{ijklm}$. Note that
\begin{align}
&\tilde{Q}_a=\sum_i|i\rangle\eta_{ia},\qquad\quad Q^a=\sum_i[i|\frac{\partial}{\partial\eta_{ia}},\nonumber\\
&\delta^{(10)}(\tilde{Q}_a) =\frac{1}{2^5}\prod_{a=1}^5 \sum_{i,j}\langle ij\rangle\eta_{ia}\eta_{ja},\qquad \quad  \delta^{(10)}(Q^a)= \frac{1}{2^5}\prod_{a=1}^5 \sum_{i,j}[ij]\frac{\partial^2}{\partial\eta_{ia}\partial\eta_{ja}}.
\end{align}

The first step is to factor the $\delta$-function, writing
\begin{equation}
\delta^{(10)}\left(\sum_{i=1}^n|i\rangle\eta_{ia}\right)=\frac{1}{\langle n-1,n\rangle^5}\delta^{(5)}\left(\sum_{i=1}^n\langle n-1,i\rangle\eta_{ia}\right)\delta^{(5)}\left(\sum_{j=1}^n\langle n,j\rangle\eta_{ja}\right).
\end{equation}
The $\delta^{(5)}$'s  provide two constraints, which we solve for  $\eta_{na}$ and $\eta_{n-1,a}$:
\begin{equation}
\eta_{n-1,a}=-\sum^{n-2}_{i=1}\frac{\langle ni\rangle}{\langle n,n-1\rangle}\eta_{ia},\qquad \eta_{na}=-\sum^{n-2}_{i=1}\frac{\langle n-1,i\rangle}{\langle n-1,n\rangle}\eta_{ia}.
\end{equation}
Since the anzatz (\ref{N5gen}) is proportional to the $\delta$-function, we  can use these expressions to simplify $P_5$,  which reduces to
\begin{equation} \label{p5c}
P_5=\frac{1}{\langle n-1,n\rangle^5}\sum_{i,j,k,l,m=1}^{n-2}c_{ijklm}\eta_{i1}\eta_{j2}\eta_{k3}\eta_{l4}\eta_{m5}.
\end{equation}
Although $c_{ijklm}$ is related to $q_{ijklm}$, we do not need the explicit relation.  $SU(5)$ R-symmetry requires that the $c_{ijklm}$ are totally symmetric in thier indices. 

Next we need to constrain $c_{ijklm}$ so that $P_5$ becomes $Q$-invariant. The condition $Q^aP_5=0$ reads, 
\begin{equation}
0=\sum^{n-2}_{i,j,k,l,m=1}[\epsilon i]c_{ijklm}\eta_{j2}\eta_{k3}\eta_{l4}\eta_{m5}=\sum_{j,k,l,m=1}^{n-2}\left[\sum_{i=1}^{n-2}[\epsilon i]c_{ijklm}\right]\eta_{j2}\eta_{k3}\eta_{l4}\eta_{m5},
\end{equation}
which leads to
\begin{equation}
\sum_{i=1}^{n-2}[\epsilon i]c_{ijklm}=0.
\end{equation}
We choose two fixed lines $i=s$ and $i=t$ and take $\epsilon=s$ or $\epsilon=t$. Then, we find
\begin{equation}
c_{tjklm}=-\sum_{i\neq s,t}\frac{[si]}{[st]}c_{ijklm},\qquad c_{sjklm}=-\sum_{i\neq s,t}\frac{[ti]}{[ts]}c_{ijklm}.
\end{equation}
We use this information to eliminate $c_{sjklm}$ and $c_{tjklm}$ in (\ref{p5c}), obtaining
\begin{align}
\langle n-1,n\rangle^5P_5=&\sum_{j,k,l,m=1}^{n-2}\sum_{i\neq s,t}^{n-2}c_{ijklm}\left(\eta_{i1}+\frac{[is]}{[st]}\eta_{t1}+\frac{[it]}{[st]}\eta_{s1}\right)\eta_{j2}\eta_{k3}\eta_{l4}\eta_{m5}\nonumber\\
=&\frac{1}{[st]}\sum_{j,k,l,m=1}^{n-2}\sum_{i\neq s,t}^{n-2}c_{ijklm}m_{ist,1}\eta_{j2}\eta_{k3}\eta_{l4}\eta_{m5}.
\end{align}
Repeating this process for  $c_{isklm}$ and $c_{itklm}$ etc.,  we find
\begin{equation}
P_5=\frac{1}{\langle n-1,n\rangle^5[st]^5}\sum_{i,j,k,l,m}c_{i,j,k,l,m}m_{ist,1}m_{jst,2}m_{kst,3}m_{lst,4}m_{mst,5}.
\end{equation}
Thus, the NMHV $n$-point superamplitude is written as
\begin{equation}
\mathcal{M}^{NMHV}_n=\sum_{i,j,k,l,m\neq s,t}^{n-2}c_{ijklm}X_{ijklm},
\end{equation}
with
\begin{align}\label{X}
X_{ijklm}=&\frac{\delta^{(10)}(\tilde{Q}_a)m_{ist,1}m_{jst,2}m_{kst,3}m_{lst,4}m_{mst,5}}{[st]^5\langle n-1,n\rangle^5},\\
m_{ist,a} = &[st]\eta_{ia} +[ti]\eta_{sa}+[is]\eta_{ta}.
\end{align}
By taking $s=n-3$, $t=n-2$, we can rewrite the superamplitude as
\begin{equation}
\mathcal{M}^{NMHV}_n=\sum_{1\leq i\leq j\leq k\leq l\leq m\leq n-4}c_{ijklm}X_{(ijklm)},\qquad X_{(ijklm)}=\sum_{P(i,j,k,l,m)}X_{ijklm}.
\end{equation}
Symmetry of the $c_{ijklm}$ has been used to order the indices, so that $P(i,j,k,l,m)$ includes all distinct arrangements for each fixed set 
$(i,j,k,l,m)$. 

Note that only $c_{11111},\,c_{11112},\dots c_{22222}$ are left in the game.  We now show that these coefficients are related to specific physical amplitudes of the basis. To simplify the discussion, we restrict to the 6-particle superamplitude and we choose $s,t = 3,4$. First we apply the 10th order Grassmann derivative  $\partial_{i,1}\partial_{i,2}\partial_{i,3}\partial_{i,4}\partial_{i,5}$ for $i=n, n-1$.  This produces the factor $\langle n,n-1\rangle^5$ which cancels the factor in the denominator of (\ref{X}).  Physically speaking, we have placed two -ve helicity gravitons at lines 5, 6.  

There remains the 5th order product of $m_{i34,a}$. We then apply the 5th order derivative $\partial_{i1}\partial_{j2}\partial_{k3}\partial_{l4}\partial_{m5}$ with the restriction $i,j,k,l,m\neq 3,4$.  The restriction means that we have placed +ve helicity gravitons at lines 3,4, and we also cancel the $1/[34]^5$  factor in (\ref{X}). There remain 6 choices of $i,j,k,l,m$,  namely $11111$ which produces a -ve helicity graviton at line 1 and leaves its +ve helicity anti-particle at line 2; then $11112$ which produces a -ve helicity gravitino at line 1 and its anti-particle at line 2; and so on until the last case $22222$ which corresponds to a +ve helicity singlet fermion at line 1 and its -ve helicity antiparticle at line 2.
In this way we arrive at the final expression for the superampliude:
\begin{align}
{\cal M}_6=& \langle-+++--\rangle X_{11111}+\langle\psi_-^{1234}\psi_-^{5}++--\rangle X_{(11112)}+\langle v_-^{123}v_+^{45}++--\rangle X_{(11122)}\nonumber\\
&+\langle\lambda_-^{12}\lambda_+^{345}++--\rangle X_{(11222)}+\langle\bar{\phi}^1\phi^{2345}++--\rangle X_{(12222)}+\langle\chi_+\chi_-++--\rangle X_{22222}.\label{NMHV}
\end{align}

With the choices  made in this derivation, that particle states from the +ve helicity wave function $\Phi$ are always at positions 2, 3, 4, while states from the -ve helicity wave function $\Psi$ are at positions 1, 5, 6.  In Sec. \ref{locality} of the text we show that superamplitude $\mathcal{M}^{NMHV}_6$ actually has full permutation symmetry in 1, 5, 6 and 2, 3, 4.  The non-manifest $1\leftrightarrow6$ and $2\leftrightarrow4$ exchange symmetries are exhibited via different but physically equivalent choices of the lines $n,n-1$ and $3,\ 4$.

\section{Examples of matrix elements of $\cN =5, L=4$ six-point superamplitude}\label{examples}
In this section, we show some example of matrix elements in \eqref{N5L4sa}.
First, we show two cases:
\begin{align}
\langle \chi_++\lambda^{123}_+v_+^{45}--\rangle=&a\langle56\rangle^4[24]^2[23](3s[12]-5[24][1|(p_2+p_3)|4\>),\\
\langle \chi_++\psi^1_+\phi^{2345}--\rangle
=&a\langle56\rangle^4[23]^3(s[12]-5[24][1|(p_2+p_3)|4\>),
\end{align}
with $s = (p_2+p_3+p_4)^2$.
In both examples, the initial expression contained a $1/[34]$ pole, and spinor manipulations using Schouten and momentum conservation were used to show that it cancels.  The result shows that graviphotons and gravitinos are present in the spectrum, although absent in the basis.

There are other examples with an initial $1/\<56\>$ pole which then cancels.  One of them is the permuted basis amplitude
\begin{align}
 \langle -\chi_-++-\chi_+\rangle
 =&5a[6|(p_3+p_4)|2\rangle[34]^4\langle15\rangle^4.\label{chiperm2}
\end{align}
Comparison with \reef{chibas} reveals a small subtlety. The two forms are properly related by the permutation $1\leftrightarrow6$ including the - sign from fermion exchange.  Another case is

We now consider two examples in which both basis amplitudes contribute with initial doubled pole $1/[34]\<56\>.$
The first one contains 6 $\chi$-fermions, $\<\bar\chi_+ \chi_-\chi_-\chi_-\bar\chi_+\bar\chi_+\>$.  This turns out to vanish after further algebra.\footnote{ It must vanish, since the only ansatz consistent with little group scaling and fermi statistics involves the combination $\<12\> \<3| +\<23\> \<1|+\<31\> \<2|$ which vanishes due to Schouten.} Indeed, we find
\bea
\<\bar\chi_+ \chi_-\chi_-\chi_-\bar\chi_+\bar\chi_+\> &=& (\pa_{234})_1(\pa_{234})_2(\pa_{234})_3(\pa_{234})_4(\pa_{234})_5 {\cal M}_6 \equiv A.
\eea
Tabulate:
\bea
(\pa_2\pa_3\pa_4)_a\d^{10}(\tQ) m_{134,b} &=& \d_{ab}\bigg(\<23\>[13]-\<24\>[41]\bigg)\\
&=& \d_{ab}[1|p_3+p_4|2\>.\label{onm134}
\eea
\bea\label{dm2}
(\pa_2\pa_3\pa_4)_a\d^{10}(\tQ) m_{234,b} &=& \d_{ab}\bigg(\<34\>[34]+\<23\>[23]+\<42\>[42]\bigg).
\eea
Then:
\bea\nonumber
&&A= \frac{5a}{[34]\<56\>}\bigg(\<34\>[34]+\<23\>[23]+\<42\>[42]\bigg)^4 \times\\
&&\bigg((p_2+p_3+p_4)^2 [1|p_3+p_4|2\>  - [1|p_3+p_4|2\> \{(\<34\>[34]+\<23\>[23]+\<42\>[42]\}\bigg).
\nonumber
\eea
The first term in the last line is the contribution of the scalar basis element and the second term is the contribution of the spinor basis element.  They cancel since $2p_j\cdot p_j = \<ij\>[ij]$ and $p_i =- |i]\<i|-|i\rangle[i|$.  

The second amplitude requires considerable manipulation before the simple final form appears:
\be
\<\bar\chi_+ \chi_-\chi_- +\bar\chi_+ -\>  = -5a ([4|p_2+p_3|6\>)^4 \<23\>[15].
\ee
 
The final example is the 6-scalaramplitude
$\langle\bar\phi^1\phi^{2345}\phi^{2345}\phi^{2345}\bar\phi^1\bar\phi^1\rangle$, which is of interest for the 
SSL issue.  We write
\begin{equation}
\langle\bar\phi^1\phi^{2345}\phi^{2345}\phi^{2345}\bar\phi^1\bar\phi^1\rangle=
(\pa_{156})_1(\pa_{234})_2(\pa_{234})_3(\pa_{234})_4(\pa_{234})_5 {\cal M}_6 \equiv B.
\ee

We calculate:
\be
(\pa_1\pa_5\pa_6)_a \d^{10}(\tQ) m_{134,b} =\d_{ab}[34]\<56\>\qquad\quad (\pa_1\pa_5\pa_6)_a \d^{10}(\tQ) m_{234,b} =0.
\ee
The  second  result means that only the scalar basis element contributes.  Using the first relation together with \reef{dm2} we find
\be
B = as(\langle23\rangle[23]+\langle24\rangle[24]+\langle34\rangle[34])^4 = a s^5,
\ee
with $s=(p_2+p_3+p_4)^2$.

We have computed more examples, for which we spare the reader.  We hope that he/she is convinced that we have performed with due diligence.

\bibliographystyle{JHEP}
\bibliography{refs}

\end{document}